\documentclass[journal,onecolumn]{IEEEtran}
\usepackage[pdftex]{graphicx}
\usepackage{url}
\usepackage{bm}
\usepackage{tabularx}
\usepackage{textcomp}
\begin{document}
	
	\title{Experimental evaluation of beamforming on UAVs in cellular systems}
	
	\author{\IEEEauthorblockN{Tomasz Izydorczyk\textsuperscript{*}, Michel Massanet Ginard\textsuperscript{*}, Simon Svendsen\textsuperscript{$\dagger$}, Gilberto Berardinelli{*} and Preben Mogensen\textsuperscript{*$\dagger$}}
	\IEEEauthorblockA{\\\textsuperscript{*}Department of Electronic Systems, Aalborg University (AAU), Denmark\\ \textsuperscript{$\dagger$}Nokia Bell Labs, Aalborg, Denmark\\ E-mail: \{ti,gb\}@es.aau.dk, mmassa18@student.aau.dk, \{simon.svendsen,preben.mogensen\}@nokia-bell-labs.com}}
	
	\maketitle
	
	\begin{abstract}
	The usage of beamforming in Unmanned Aerial Vehicles~(UAVs) has the potential of significantly improving the air-to-ground link quality. This paper presents the outcome of experimental trial of such a UAV-based beamforming system over live cellular networks. A testbed with directional antennas has been built for the experiments. It is shown that beamforming can extend the signal coverage due to antenna gain, as well as spatially reduce interference leading to higher signal quality. Moreover, it has a positive impact on the mobility performance of a flying UAV by reducing handover occurrences. It is also discussed, in which situations beamforming should translate into the uplink throughput gain. 
	\end{abstract}	

	\section{Introduction}	
	
	Unmanned Aerial Vehicles~(UAVs) will require a reliable and high uplink throughput communication link to ensure flight's safety and serve foreseen use cases such as package delivery or surveillance~\cite{bigsurvey2}. For example, up to 50~Mbps of continuous uplink throughput will be required to send uncompressed video from a UAV to a cloud for machine learning image processing~\cite{bigsurvey}.
	
	Cellular systems including Long Term Evolution~(LTE) are a potential candidate to cope with the stringent requirements and provide worldwide connectivity to UAVs. However increased Line of Sight~(LoS) probability towards interfering base stations due to lack of obstructions in the Air-to-Ground~(A2G) channel results in a Signal-to-Interference plus Noise Ratio~(SINR) reduction~\cite{studyit}. In the downlink, the interference will result in lower service reliability for a UAV, while higher uplink interference radiated by a UAV to the network will result in decreased performance of other incumbents of the cellular systems, such as ground users or other UAVs.
	
	There are many proposed solutions on how to deal with interference while maintaining UAV connectivity. In~\cite{3Dbeam1} and~\cite{3Dbeam2} authors discuss how massive Multiple Input Multiple Output~(MIMO) and 3D beamforming at the base stations can be utilized for UAV connectivity. In~\cite{NOMA} and~\cite{cell_free} authors claim that Non-Orthogonal Multiple Access~(NOMA) and cell-free massive MIMO can outperform 3D beamforming and provide even better UAV connectivity. Finally, there are multiple~(\cite{planning2},~\cite{planning}) attempts to optimize the path of a flying UAV in order to contain the radiated/absorbed interference. 
	
	Use of multiple antenna techniques at a UAV, such as beamforming, is yet another foreseen solution which in addition does not require any network hardware changes. Simulations in~\cite{beam} showed the potential of UAV-side beamforming to boost signal strength while spatially reducing interference towards unwanted directions. However this benefits were only shown in simulation and no experimental validation is available in the literature.
	
	Practical implementation of a beamforming system is a non-trivial task due to space/weight/power constraints of the UAV. Use of directional antennas is one possible simplification of beamforming. Although it reduces the arbitrary number of directions to the number of antennas, it facilitates the development time and cost of the hardware platform, accelerating research activities.
	
	In this work, a complete design of a flexible UAV testbed using directional antennas for beamforming evaluation is presented and further used in a measurement campaign to understand the promises of uplink directional communication for cellular-connected UAVs. The contributions of this paper are twofold. First, to the best of the authors knowledge, it is the first paper explaining in detail the hardware implementation of a UAV-based directional antenna's switching system. Second, this paper contains the measurement results of a first trial of beamforming-enabled UAV flying over real cellular networks, discussing insights and potential benefits of this technology. 
	
	The rest of the paper is structured as follows. In Section~\ref{s2} the high-level system design is presented together with the remarks on the flexibility aspects of the platform. It is followed by Section~\ref{s3.5} focused on the implemented beam steering algorithm. Section~\ref{s4} describes the results of a measurement campaign conducted using live cellular networks. The discussion on potential benefits of uplink beamforming is made in Section~\ref{s5}.

	\section{System Design}
	\label{s2} 	
\begin{figure}[!t]
		\centering	
		\includegraphics[width=\linewidth]{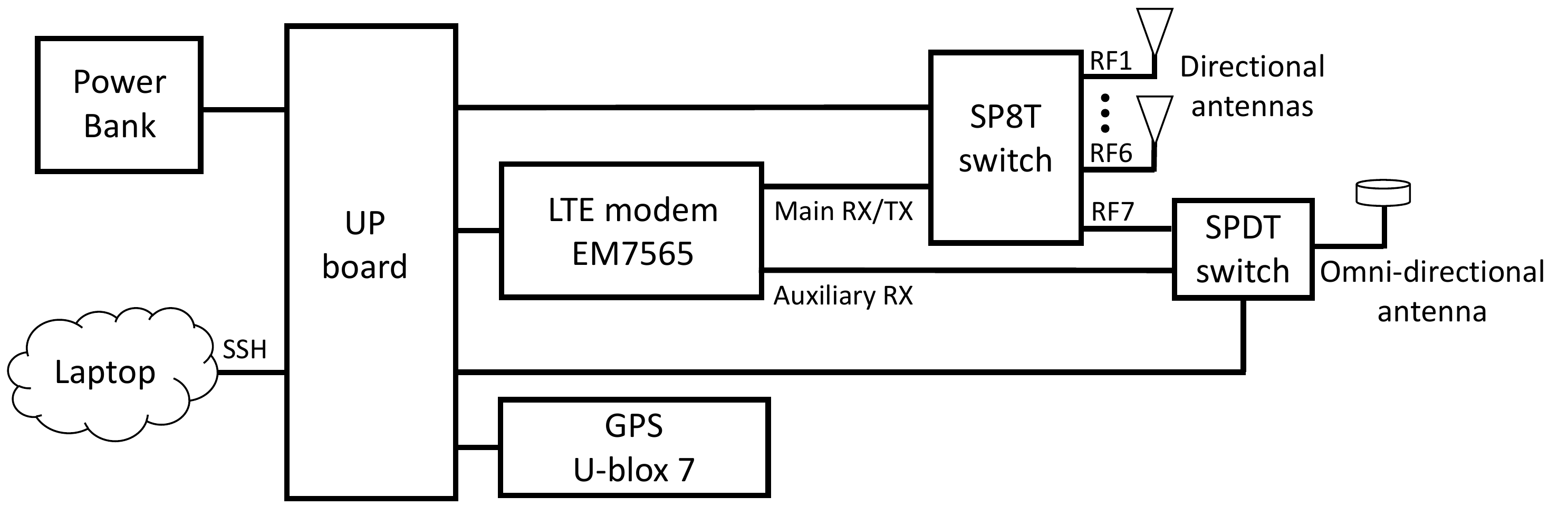}
		\caption{Schematic of the designed testbed}
		\label{Scheme_equipment}
	\end{figure}

	The proposed system was designed to operate using live cellular networks with the target of being used for a number of different research activities from Path Loss~(PL) measurements and interference characterization up to uplink throughput studies. The system as presented in Figure~\ref{Scheme_equipment} is composed of an antenna array, consisting of six directional patch antennas and a monopole omni-directional antenna. Two Radio Frequency~(RF) switches, an embedded computer and a LTE modem complement the design. When flying, the system can either operate autonomously with the mission (trajectory, switching algorithm, data transmission etc.) programmed on the embedded computer or directly controlled from the ground using Secure Shell~(SSH) connection. This design was initially inspired by~\cite{switched1}, where an array of directional antennas was used to establish WiFi connectivity.
	
\begin{figure} [!t]
		\centering	
		\includegraphics[width=0.8\linewidth]{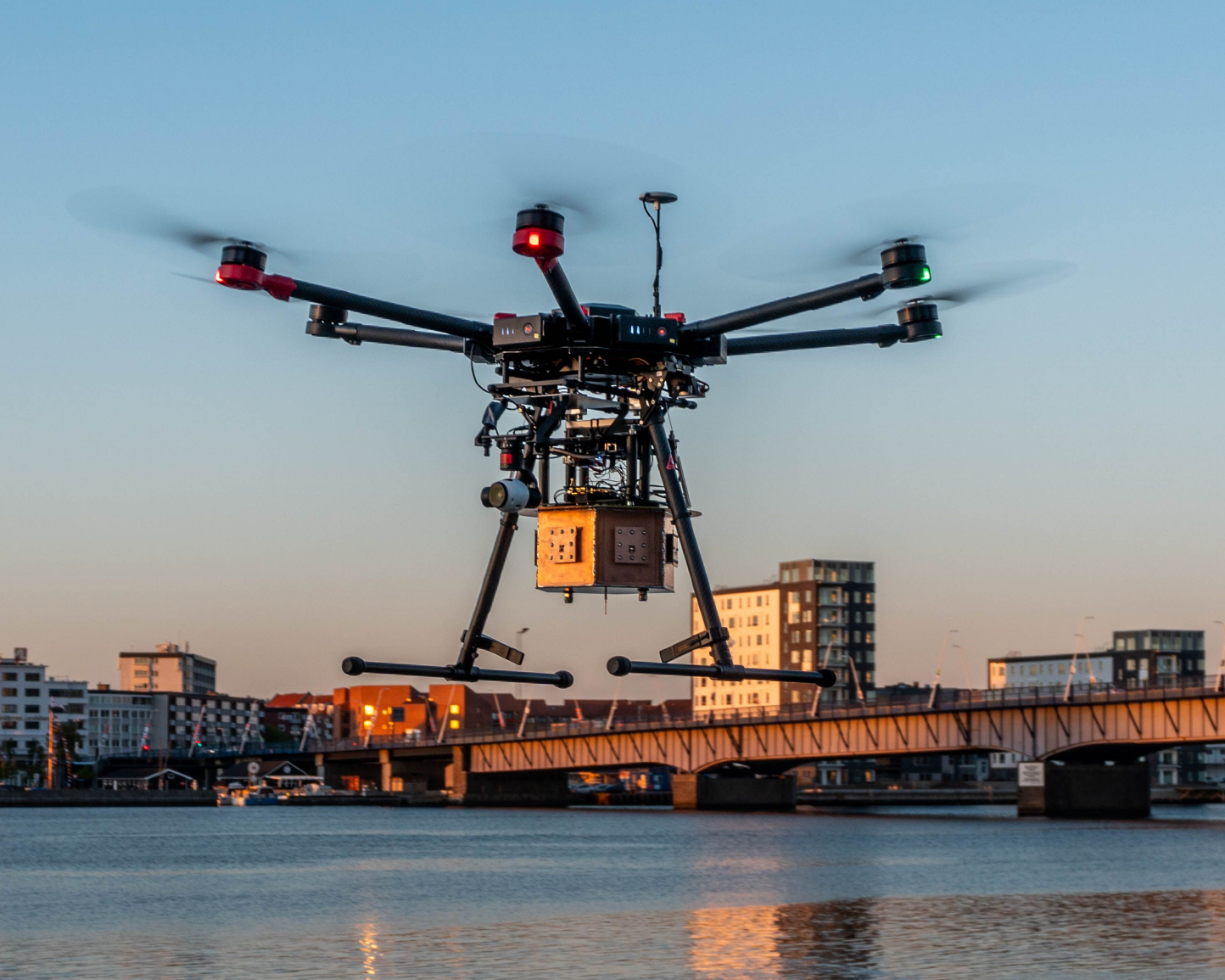}
		\caption{Assembled setup on a UAV}
		\label{drone_equipment}
	\end{figure}

	The testbed was mounted using a customized carbon fiber structure on a DJI M600 drone. Its total weight is approximately 2~kg, allowing for more than~35~minutes of continuous flight time. It can be powered using a simple power bank as all components accept a standard 5V power source. The final assembled setup is shown in Figure~\ref{drone_equipment}.
	
  	Two RF Switches are used. Single Pole 8 Throw~(SP8T) switch connects all antennas (omni and directional) to the main antenna port of the LTE modem. Single Pole Double Throw~(SPDT) switch connects the omni-directional antenna to its auxiliary port. Switches are controlled by the embedded computer built based on Intel's \emph{UP board}. 
  
 	The cellular modem used in the platform is a \emph{Sierra Wireless EM7565}. Please note that the modem is not only used to provide the network (and SSH) connectivity, but it is also used to provide valuable network information which can be used for research purposes. These include among others measurement reports from up to eight cells, quantified as Reference Signals Received Power~(RSRP), Reference Signal Received Quality~(RSRQ), Received Signal Strength Indicator~(RSSI) and SINR. Moreover the information of the serving cell is accessible and can be used, for example to quantify a number of network handovers. 
  
  	In this work, the modem is set to operate in LTE-only mode in the desired band~3~(1.7~GHz uplink and 1.8~GHz downlink frequencies). The auxiliary antenna port can be used as a receive diversity port as explained later in Section~\ref{s3.5}. It can be activated within a modem using AT commands. GPS information is acquired from the \emph{Ublox-7} GPS module connected to the embedded computer and is used to provide real-time location and heading orientation to the system.
  	 	
	\subsection{Antenna design}
	\label{s3}	
	
\begin{figure}
		\centering	
		\includegraphics[width=\linewidth]{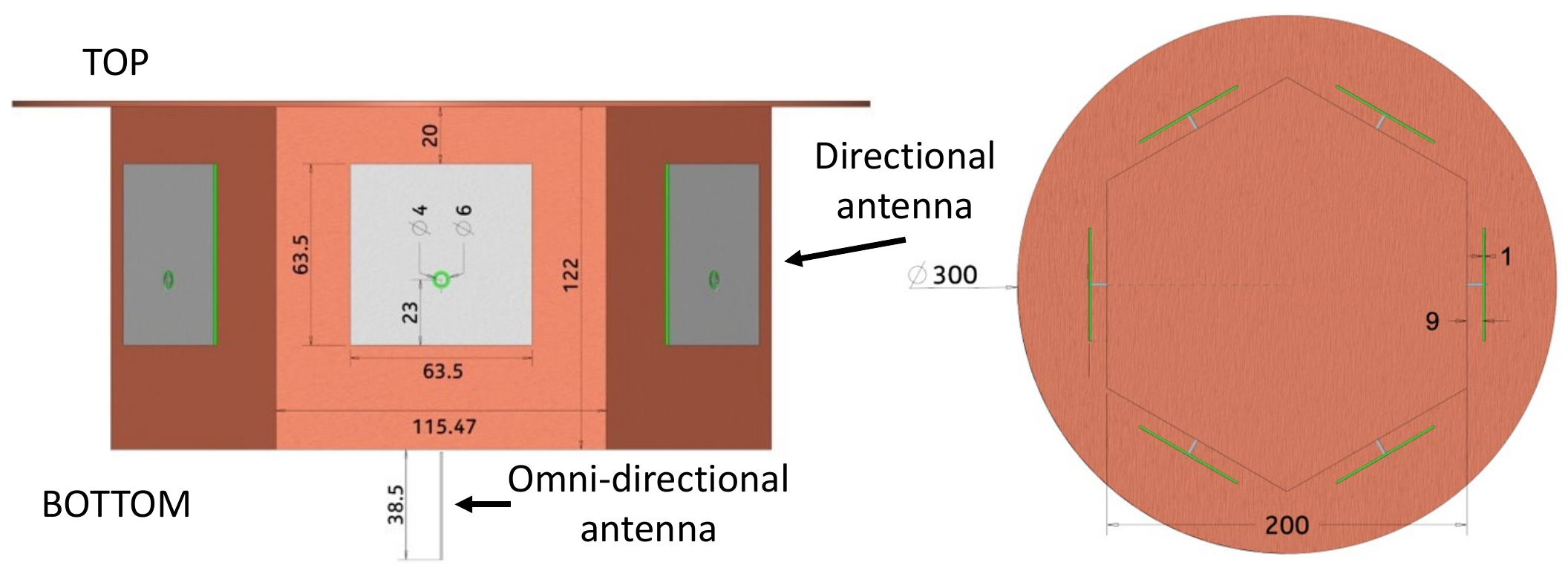}
		\caption{Designed antenna array side and top view}
		\label{antenna_side}
	\end{figure}

\begin{figure}
		\centering	
		\includegraphics[width=\linewidth]{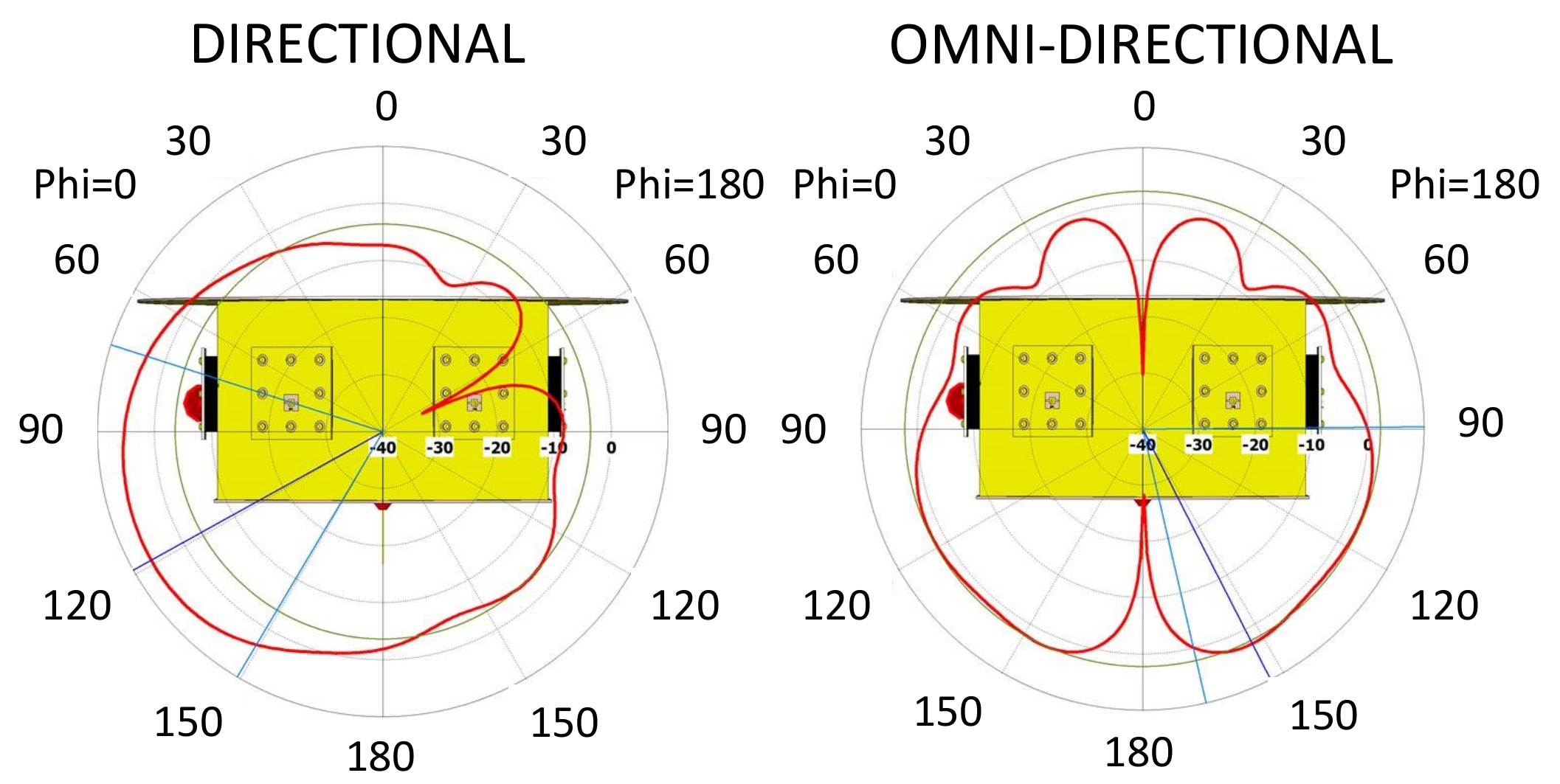}
		\caption{Simulated far field antenna pattern for both antenna types}
		\label{beamwidth_patch}
	\end{figure}

	The antenna array is composed of six identical patch antennas deployed on a hexagonal cylinder-like structure. In this way patch antennas are equally spaced and pointing at different directions, as shown in Figure~\ref{antenna_side}. Figure~\ref{beamwidth_patch} (left side) presents the simulated radiation pattern of one directional antenna using CST studio with the designed Half Power Beamwidth~(HPBW) of approximately $70^o$ giving a desired, slight overlap between different patches. The elevation angle is approximately $61^o$. The directivity of the final design is 6.9~dB with a realized gain of 6.4~dB. The bigger ground plane located behind each patch antenna helps to increase the front to back ratio to above 15~dB and the Side Lobe Level to -10~dB.
	
\begin{figure}
		\centering	
		\includegraphics[width=\linewidth]{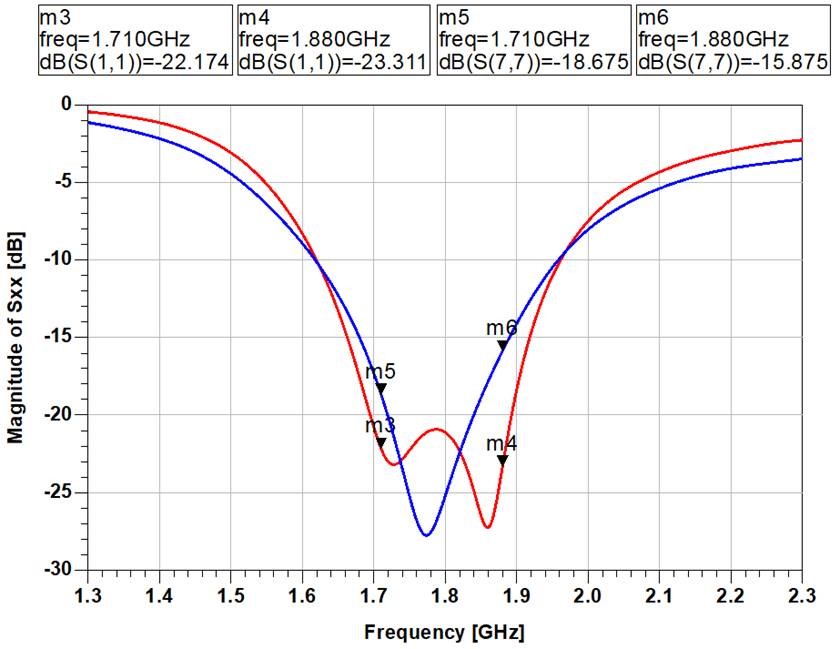}
		\caption{Simulated S11 parameters for patch antenna~(red) and monopole~(blue)}
		\label{s_parameters}
	\end{figure}

	The omni-directional antenna located at the ground plane at the bottom of the cylinder is a simple monopole. It provides approximately~2~dB gain as shown on the right side of Figure~\ref{beamwidth_patch}. The simulated S11 parameters of both antennas are shown in Figure~\ref{s_parameters}. Only one patch antenna is shown as the others are considered to be identical. As shown, the designed antennas cover the frequency range from 1.7~GHz to 1.9~GHz (with the magnitude of S11 below -15~dB) where both uplink and downlink of LTE band~3 are located. All antennas are connected with the RF switches using the SMA cables of the same length placed inside the designed structure.
		
	\subsection{Design flexibility}
	UAV measurements are usually costly and time consuming, therefore flexibility and versatility of the measurement platform are desired. The modular design of the testbed facilitates changes of only single elements in case different frequency bands or antenna configurations are to be studied. This can greatly reduce the development time since most of the hardware elements can be reused. 
	
	Another example of flexible design is the addition of a monopole omni-directional antenna. Very often experimental results of beamforming are to be compared with a reference antenna. By adding this antenna at the design phase, the potential set of measurements can be maximized without the need to exchange the hardware or even land the UAV.

	\section{Beam switching algorithm and mobility management}
	\label{s3.5}
	The beam switching algorithm implemented in this testbed relies on three key information. Serving Cell ID provided by the LTE modem, together with the known a priori list of GPS coordinates of Base Stations (BSs) are used to identify the location of a serving cell. By knowing the instantaneous GPS location and heading of the UAV, the antenna which points towards the direction of the desired BS is selected. In the LoS-dominant scenarios, often observed in cellular-connected UAVs, this kind of simple antenna-selection algorithm can be used without the need for real-time beam training techniques. 
	
\begin{figure}
		\centering	
		\includegraphics[width=0.8\linewidth]{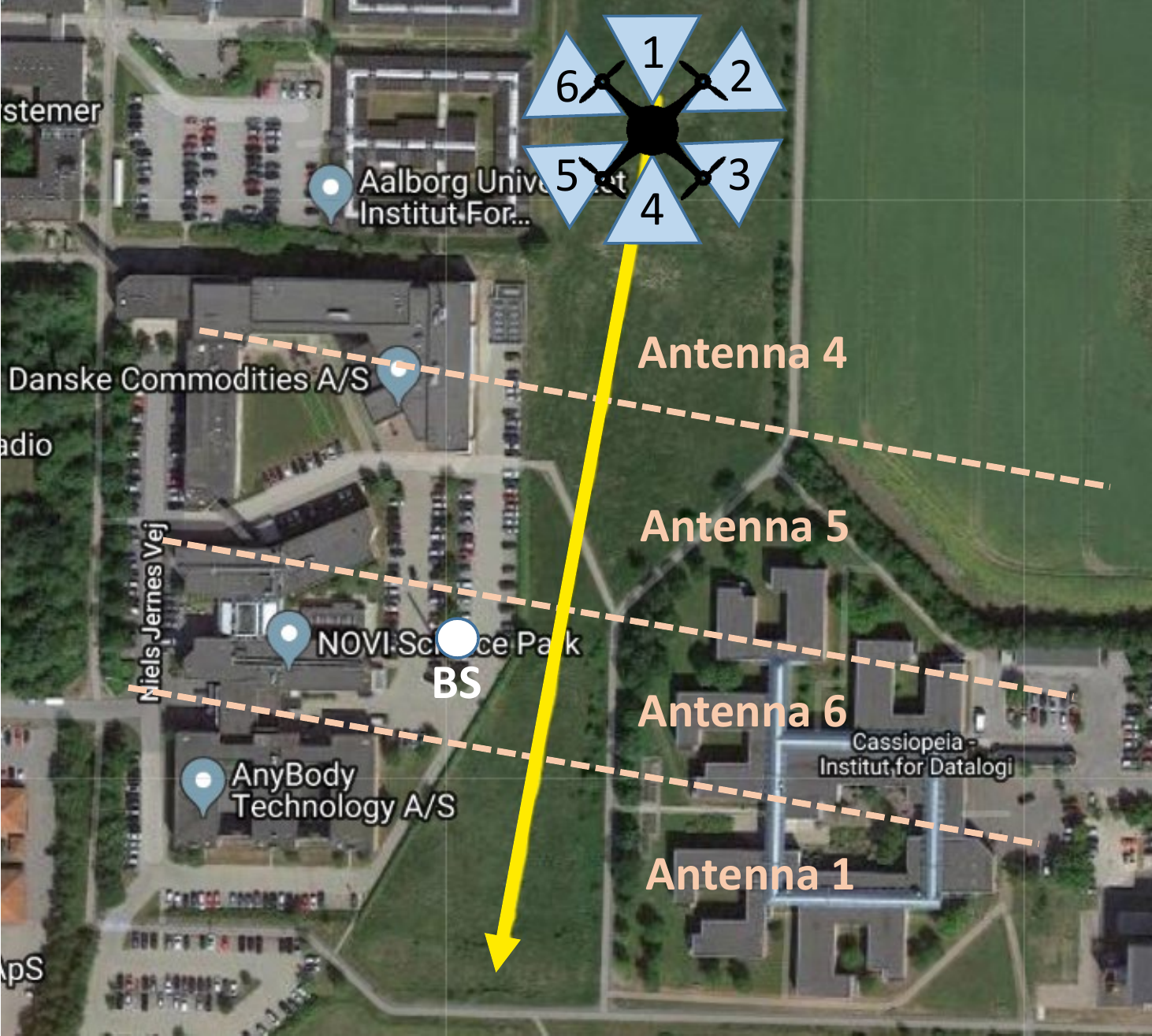}
		\caption{Beam switching example with UAV flight path (yellow) and the switching antenna regions}
		\label{map_switching}
	\end{figure}

	Figure~\ref{map_switching} presents the execution of the beam switching algorithm during one of the preliminary validation measurements. During the flight, the UAV approached the serving BS whose position is highlighted in the figure, flying over a straight trajectory. The UAV kept constant heading towards north. Real time computed antenna index was saved for offline analysis. While flying, the chosen antenna changed four times as illustrated in the figure.
	
	Directional antennas may affect the mobility management protocols. In cellular systems, the serving cell of a user is selected based on highest measured and reported RSRP by the modem. The directional antenna pattern would enhance/mitigate signals from neighbor cells located at specific directions leading to potentially suboptimal cell selection. The proposed testbed offers the flexibility to mitigate the occurrences of wrong cell selection by activating the omni-directional antenna on the auxiliary receiving port. In this configuration, the modem reports to the network the highest RSRP value received from the two different antenna ports. The UE can then measure cells over all directions, thus mitigating the effect of the nulls of the directional antenna while preserving its gain in the beam direction.

	\section{Measurement campaign}
	\label{s4}		
	\subsection{Measurement scenario}
	The target of the measurement campaign was to understand the impact of beamforming on a UAV's uplink performance when connected to the cellular networks. The measurement campaign took place in the city center of Aalborg, Denmark. Full buffer uplink transmission was implemented in the measurement system, in which a large file was uploaded to a local server. To isolate the effect of beamforming from the varied network load, the measurements were taken during night hours, when limited activity of the ground users was experienced.
	
\begin{figure}
		\centering	
		\includegraphics[width=\linewidth]{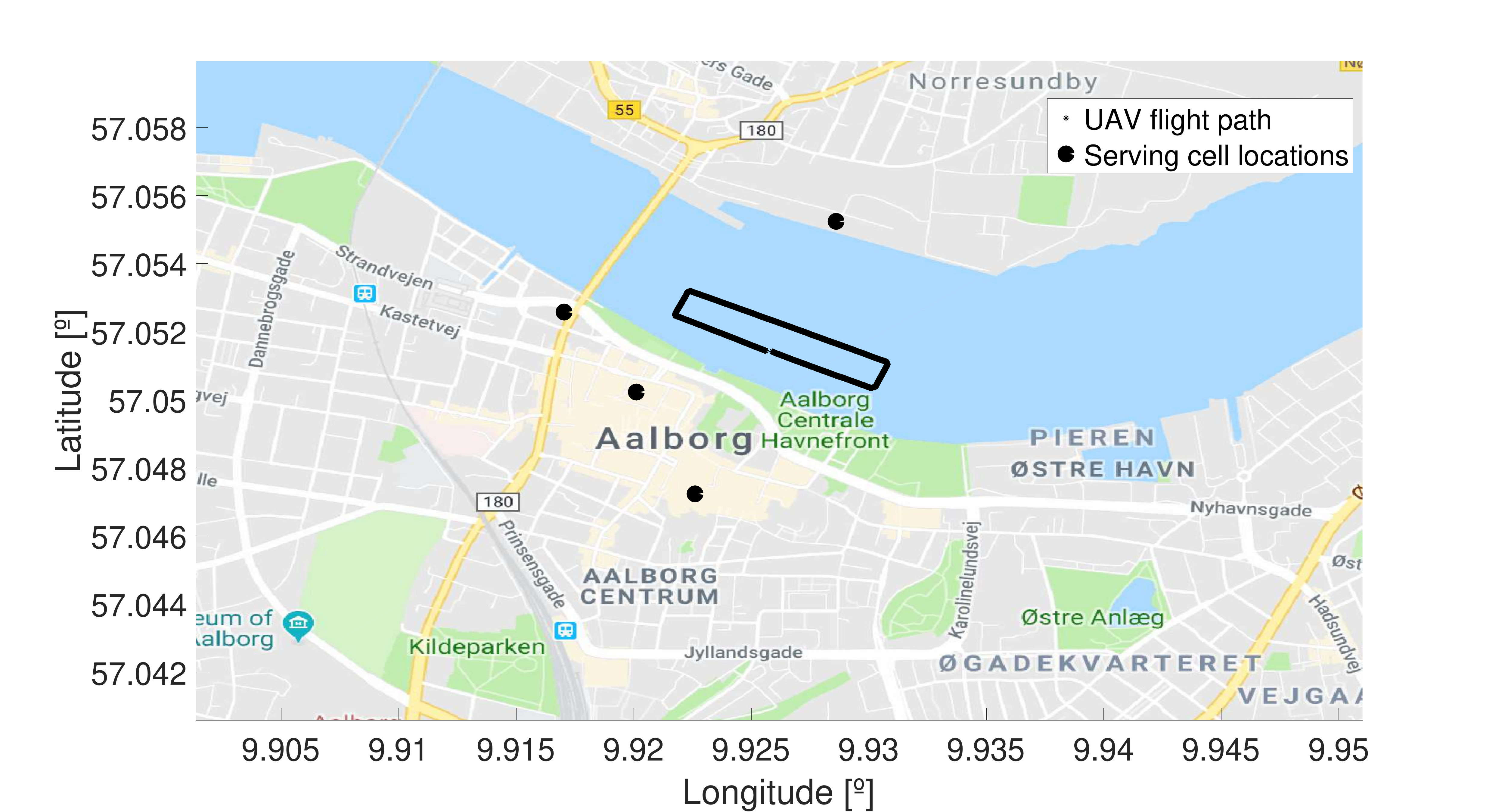}
		\caption{Executed flight path}
		\label{flight_path}
	\end{figure}

	Two UAV flights were conducted over the same path shown in Figure~\ref{flight_path}. In the first flight, the UAV was using the real-time chosen directional antenna, together with omni-directional antenna activated on the auxiliary receiving port. During the second flight, only the omni-directional monopole antenna was used connected to the main antenna port. Both flights were performed during one mission without landing the UAV. The mission was repeated for two different heights: 10~m (imitating a ground-level scenario) and 40~m (maximum height allowed by the regulations due to airport proximity). UAV flew with a constant speed of 10~km/h and throughout all flights its heading was kept constant towards north regardless of the moving direction. To facilitate understanding, in Figure~\ref{flight_path}, black dots represent the location of serving cells observed during measurements. Each dot represent a cell tower with a three sectors (cells) of a mobile network.
	
	\subsection{Results}	
	While flying, every 200~ms the LTE modem installed on a UAV reported some LTE metrics measured with the selected antenna. First, downlink RSRP and RSRQ are studied to show the effect of antenna directionality. 
	
\begin{figure}
		\centering	
		\includegraphics[width=\linewidth]{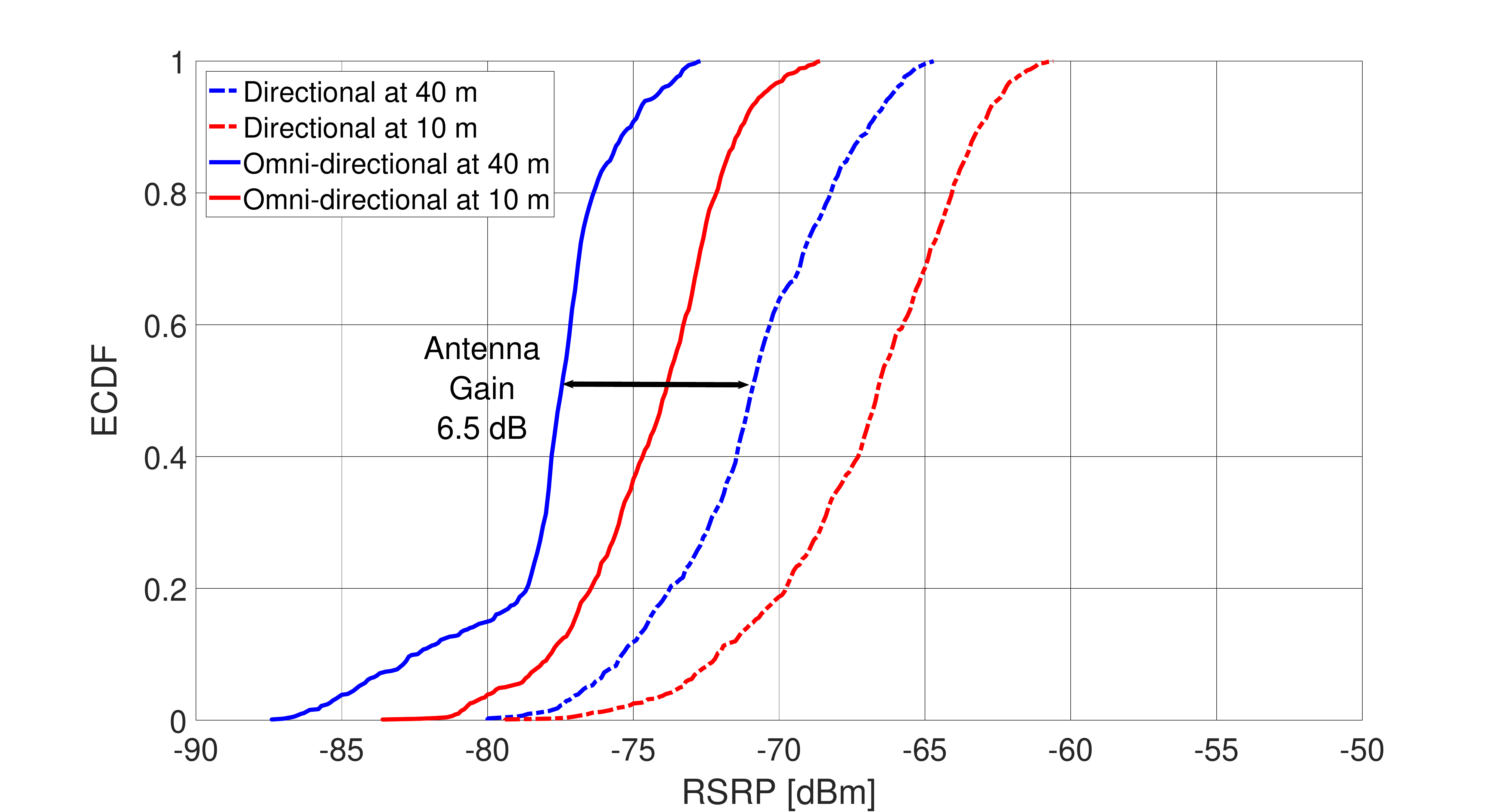}
		\caption{ECDF of the measured RSRP of the serving BS}
		\label{CDF_RSRP}
	\end{figure}

	The Empirical Cumulative Distributed Function~(ECDF) of the recorded downlink RSRP for both antenna systems is presented in Figure~\ref{CDF_RSRP}. A continuous RSRP gain by the directional antennas with respect to the omni-directional case can be observed. This, together with median RSRP gains of approximately 6.5~dB, corresponds to the directional antenna gain, therefore validating the proper execution of the beam switching algorithm. In general observed values of RSRP decreased with height proving that the deployed cellular systems are optimized for the ground communication.
	
\begin{figure}
		\centering	
		\includegraphics[width=\linewidth]{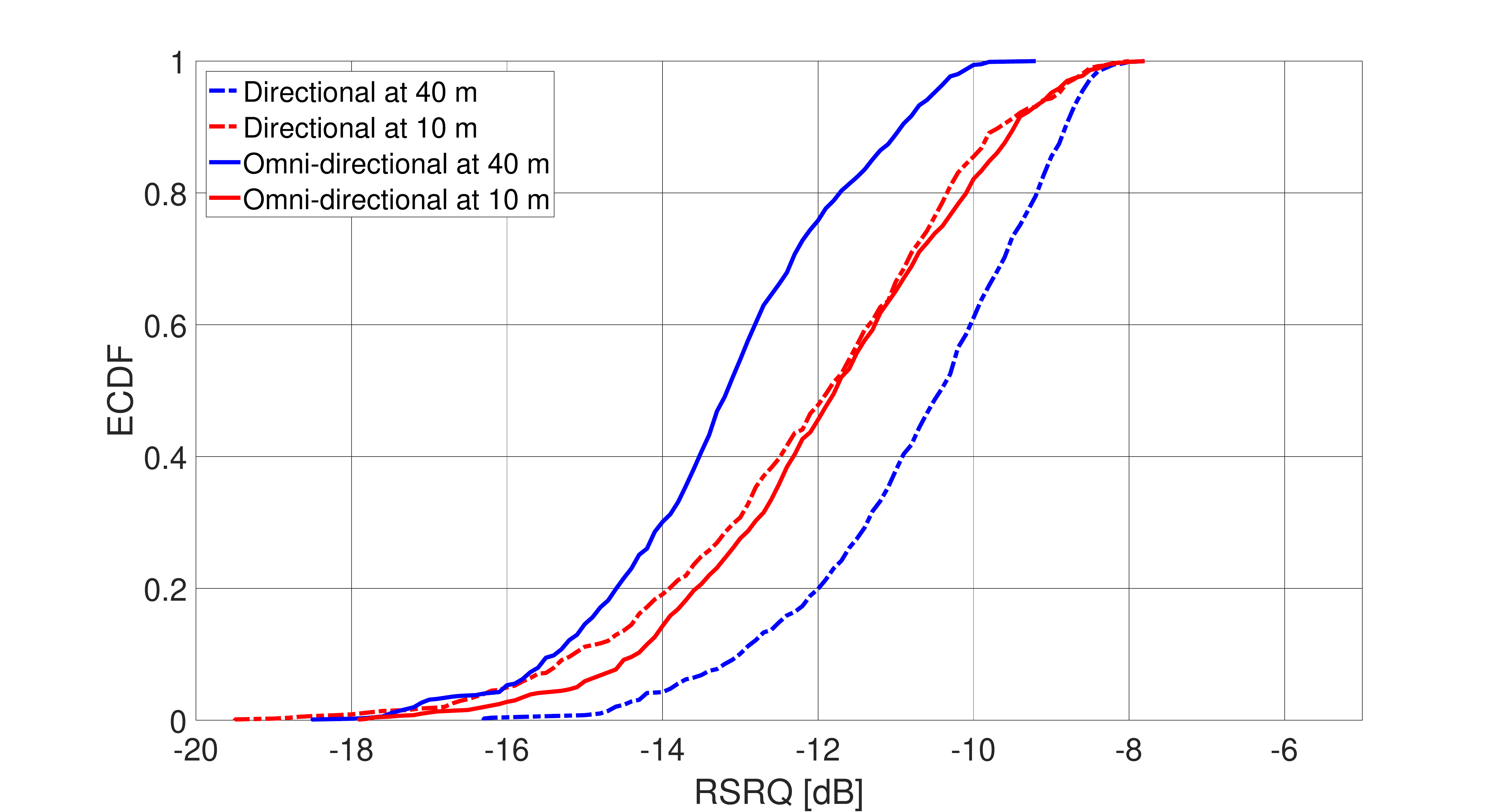}
		\caption{ECDF of the measured RSRQ of the serving BS}
		\label{CDF_RSRQ}
	\end{figure}

	Second studied metric is downlink RSRQ as presented in Figure~\ref{CDF_RSRQ}. The shown results lead to an interesting remark. There is no gain in RSRQ for directional antenna system if a UAV is flying at 10~m while there is a substantial gain observed at 40~m. The reason for this is the increased benefit of spatial interference mitigation observed at higher heights. When flying above rooftops, strong interfering signals from a large number of cells become visible leading to RSRQ decrease when omni-directional antenna is used. Conversely, due to spatial interference filtering when UAV is using a directional antenna, the RSRQ values increased leading to a 3~dB gain. Higher RSRQ usually leads to higher order modulation and coding schemes being used at the transmitter and therefore translates into improved data rates. At the low heights, the impact of interference is negligible and observed RSRQ values are similar regardless of the antenna system.
		
\begin{figure}
		\centering	
		\includegraphics[width=\linewidth]{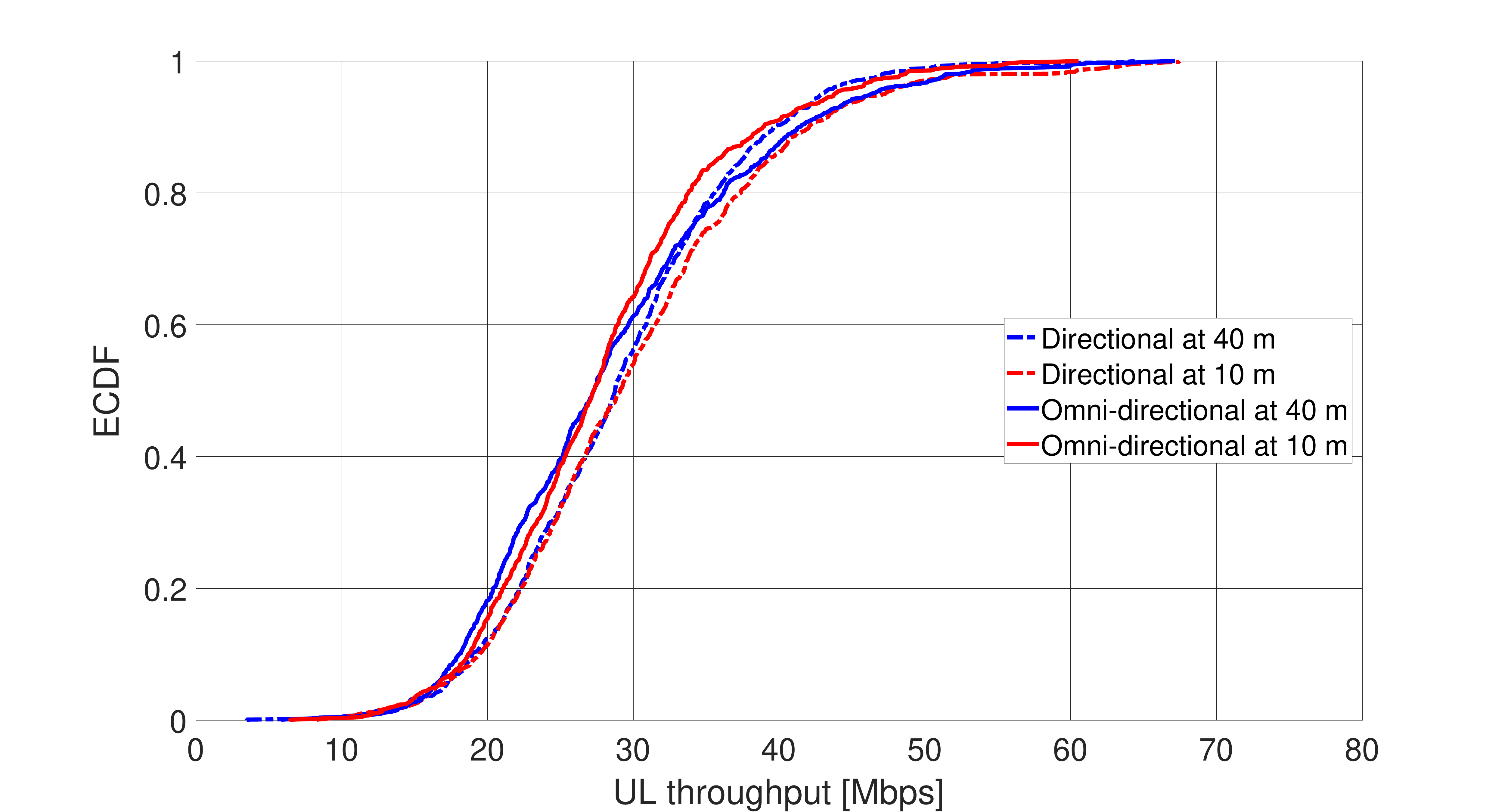}
		\caption{ECDF of the measured uplink throughput of the serving BS}
		\label{CDF_TPUT}
	\end{figure}

	After studying downlink metrics, next in Figure~\ref{CDF_TPUT}, the ECDF of uplink throughput is considered. Not surprisingly the results follow the ones observed in~\cite{moj_globecom}, where similar study was conducted for vehicular communication. As the entire flight was carried in the high RSRP regime, due to uplink power control operating in the linear region, the directional antenna gain was compensated by the network issuing lower uplink transmit power in the directional antenna case. At a first glance, the shown results indicate, that there is no user-centric benefit of using high directional antennas at the UAV. However, as discussed later, the potential benefits will become visible if UAV flies in the low RSRP regime or when number of UAVs in the same region is increased.

\begin{figure}
		\centering	
		\includegraphics[width=\linewidth]{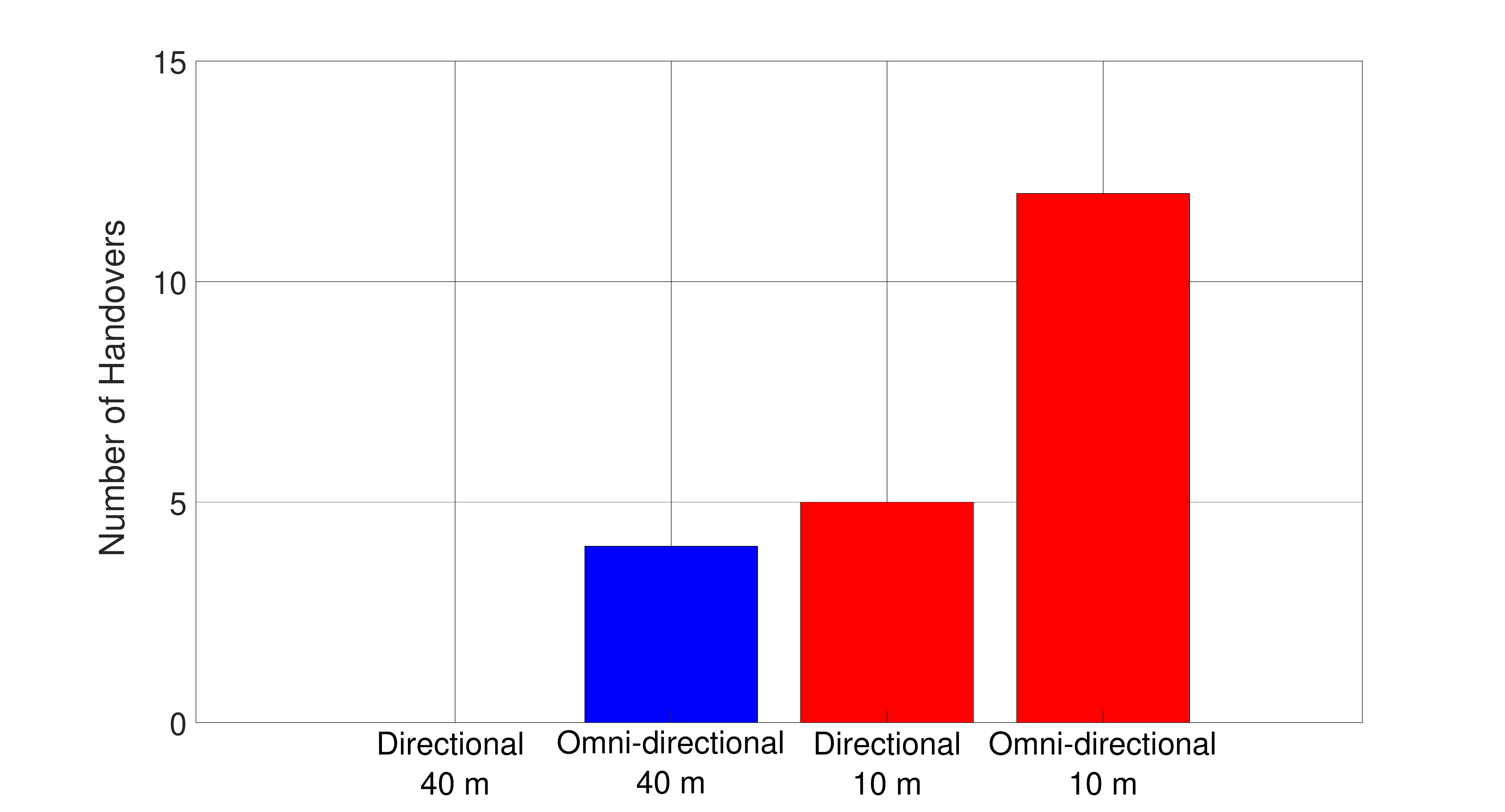}
		\caption{Number of handovers observed during measurements}
		\label{n_HOs}
	\end{figure}

	Finally, the impact of directional antennas on the network mobility is studied in Figure~\ref{n_HOs}, where total number of handovers per flight is presented for all four flights. Here, one can observe that, when using directional antennas, the total number of handovers can be reduced with respect to a UAV flying at the same height with an omni-directional antenna. This is due to the capability of the UAV to maintain connectivity to the same cell by pointing the antenna towards its direction. At 10~m the total number of handovers was reduced from 12 to 5, while at 40~m, there were no handovers when using directional antennas, as the UAV kept connectivity to the same cell throughout the entire flight!

	\section{Conclusions and Discussion}
	\label{s5}		
	In this work, a multi-functional UAV platform for beamforming experimentation is presented and used in a measurement campaign using live LTE network. Although using only six antennas (directions) may seem as a limitation when studying beamforming performance, it is worth to note that one can steer the UAV in an arbitrary direction and therefore point the beam towards all the possible angles, as UAV's heading information is acquired from the GPS.
		
	\subsection{Uplink throughput improvements}
	Presented measurement results indicate that even when equipped with beamforming system, a UAV is not benefiting from uplink throughput improvement due to uplink power control compensating the higher antenna gain. It is worth to note that such gains are instead expected to be observed in the low coverage scenario (as cell edge or even network edge - as seashore). In such cases, a UAV will transmit with maximum power and uplink power control will not be able to fully compensate the antenna gain. This will lead to beamforming gain translating to coverage extension and thus relative throughput improvement. 
	
	The conducted measurements consisted of only a single flying UAV. It is expected that with more UAVs, their radiated uplink interference will impact the observed uplink throughput. As a potential future work, we recommend conducting experiments or simulations showcasing the impact on the network performance of multiple UAVs equipped with beamforming capabilities. Our hypothesis is that, the spatially shaped interference would not only result in the improved downlink RSRQ of each UAV, but will also result in improved uplink SINR levels at the base stations (with respect to UAVs with omni-directional antennas) leading to higher uplink throughput.
	
	\subsection{Beamforming impact on mobility}
	As presented in the previous section, beamforming can help reducing the number of handovers due to longer connection to the same serving cell. This may be seen both as positive (less handovers means better reliability and lower latency since service interruptions during handover time are avoided) but also as negative (longer connectivity can potentially result in better cell being disregarded). In this measurements a naive beamforming strategy was used. It is expected, that when using other beamforming strategies, as for example interference aware RSRQ-based beam switch, beamforming can be used to steer the connectivity towards less interfered directions, possibly resulting in improved network performance.
		
	\bibliographystyle{ieeetr}
	\bibliography{references}
\end{document}